\documentclass[twocolumn,preprintnumbers,amsmath,amssymb,pra]{revtex4}
\usepackage{graphicx}
\usepackage{dcolumn}
\usepackage{bm}
\usepackage{times}
\usepackage[colorlinks,citecolor=blue,linkcolor=red]{hyperref}
\usepackage{color}
\usepackage{comment}

\begin{document}

\title{Strong system-bath coupling induces negative differential thermal conductance and heat amplification in nonequilibrium two-qubits systems}

\author{Huan Liu$^{1}$}
\author{Chen Wang$^{1,}$}\email{wangchenyifang@gmail.com}
\author{Lu-Qing Wang$^{2}$}
\author{Jie Ren$^{2,}$}\email{Xonics@tongji.edu.cn}
\address{
$^{1}$Department of Physics, Zhejiang Normal University, Jinhua 321004, Zhejiang , P. R. China\\
$^{2}$Center for Phononics and Thermal Energy Science, China-EU Joint Center for Nanophononics, \\
Shanghai Key Laboratory of Special Artificial Microstructure Materials and Technology,  \\
School of Physics Sciences and Engineering, Tongji University, Shanghai 200092, China
}

\date{\today}

\begin{abstract}
Quantum heat transfer is analyzed in nonequilibrium two-qubits systems
by applying the nonequilibrium polaron-transformed Redfield equation combined with full counting statistics.
Steady state heat currents with weak and strong qubit-bath couplings are clearly unified.
Within the two-terminal setup, the negative differential thermal conductance is unraveled with strong qubit-bath coupling and finite qubit splitting energy.
The partially strong spin-boson interaction is sufficient to show the negative differential thermal conductance.
Based on the three-terminal setup, that two-qubits are asymmetrically coupled to three thermal baths, a giant heat amplification factor is observed with strong qubit-bath coupling.
Moreover, the strong interaction of either the left or right spin-boson coupling is able to exhibit the apparent heat amplification effect.
\end{abstract}

%\pacs{}
%keywords

\maketitle

\section{Introduction}

Understanding mechanism and efficient manipulation of heat energy at nanoscale is
of fundamental significance in the scientific community~\cite{slepri2003pr,gchen2005book,ydubi2011rmp,nbli2012rmp,lee2013nature}.
One long-standing goal is to establish basic units of functional thermal devices
for the implementation of heat transport and information processing~\cite{bli2006prl,lwang2007prl,lwang2008prl,kstan2012prl,jren2015aip}.
Many theoretical models have been proposed for thermal operations, ranging from the thermal diode~\cite{bli2004prl,lfzhang2009prb,afornieri2014apl,rsanchez2015njp}, thermal transistor~\cite{bli2006prl,tojanen2008prl,truokola2011prb,pba2014prl,jhjiang2015prb,kjoulain2016prl,afornieri2016prb,rsanchez2017prb,gcraven2017prl,yczhang2017apl,yczhang2018epl}, thermal memory~\cite{lwang2007prl,vkubytskyi2014prl,cguarcello2018pra} and even thermal computer~\cite{lwang2008prl}.
And tremendous experiments were conducted to
realize these novel thermal operations~\cite{cwchang2006science,fgiazotto2012nature,syigen2014nl,kjoulain2015apl,bdutta2017prl,dzhao2017nc}.
%2. Heat energy can be transferred in several types.
In particular for the thermal transistor, negative differential thermal conductance (NDTC) and heat amplification
are considered as two key ingredients~\cite{nbli2012rmp,dhhe2009prb,dhhe2010pre,dhhe2014pre}.

The NDTC was originally exploited in nonlinear phononic lattices coupled to two terminals by B. Li and his colleges~\cite{bli2006prl},
where the heat current shows suppression by the increase of the temperature bias between thermal baths.
Based on the NDTC, the heat amplification was also exploited within the three-terminal phononic setup.
Consequently, these concepts have been extensively investigated in various quantum thermal transistors.
Particularly for the quantum dots, the capacitively Coulomb interaction allows the electronic fluctuation
to transfer heat, and the asymmetric Coulomb blockade is crucial for the NDTC~\cite{truokola2011prb,yczhang2017apl}.
For the Josephson junctions, the coherent phase modulates heat transport between superconductors,
and the analogous singularity-matching-peak effect originates the NDTC and heat amplification~\cite{afornieri2016prb,jren2013prb11,afornieri2017nnano}.
While for the near-field thermal transistor, a tiny change around the critical temperature of the insulator-metal transition material
dramatically reduces the flux into the drain~\cite{pba2014prl}.
%For the magnonic hybrid systems,

Recently, the quantum thermal transistor was theoretically unravelled in the nonequilibrium spin-boson models~\cite{agleggett1987rmp,dsegal2005prl,uweiss2008book}.
Specifically, in the qubit-qutrit system, proper energy of the qutrit is crucial for the realization of
thermal transistor effect~\cite{bqguo2018pre}.
In resonant three-qubits systems, the NDTC and giant heat amplification factor were observed
both in weak and strong coupling regimes~\cite{kjoulain2016prl,cwang2018pra}, which are attributed to the three-terminal cooperative processes.
Moreover, the strong spin-boson interaction is unravelled to be crucial to exhibit these far-from equilibrium features~\cite{cwang2018pra}.
While for the nonequilibrium two-qubits system, in which the steady state heat transport has been preliminarily investigated~\cite{lawu2011pra,akato2015jcp},
the NDTC and heat amplification is still lack of exploration, particularly in strong system-bath coupling regime.
Based on the two-qubits system we raise questions:
\emph{(i) can the NDTC be found with strong qubit-bath coupling within the two-bath setup?
(ii) what is the role of the strong coupling on the NDTC?
Moreover, considering the analysis of heat flow in the two-qubit system asymmetrically coupled to three-thermal baths~\cite{zxman2016pre},
can we observe heat amplification effect with strong qubit-bath coupling?}

\begin{figure}[tbp]
\begin{center}
\hspace{-3.0cm}
\includegraphics[scale=0.38]{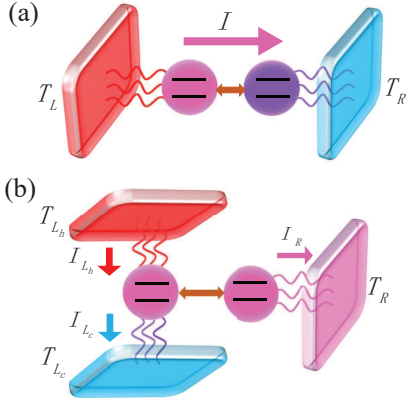}
\vspace{-1.0cm}
\end{center}
\caption{(Color online) Schematic illustration of quantum thermal transfer in two-qubit systems driven by
(a) two thermal baths (square panels with temperatures $T_L$ and $T_R$) with $I$ the heat flow and
(b) three thermal baths (square panels with temperatures $T_{L_h}$, $T_{L_c}$ and $T_R$) with $I_{L_{h(c)}}$ and $I_R$ heat currents.
The circles describe two-level qubits with horizontal solid lines
the qubit energy levels, the wave-like curves show the interaction between thermal baths and qubits system, and the double-arrowed brown lines
describe the inter-qubit coupling.
}~\label{fig1}
\end{figure}

In the present paper, we investigate steady state quantum heat transfer in nonequilibrium two-qubits systems in Fig.~\ref{fig1},
by applying the nonequilibrium polaron-transformed-Redfield equation (NE-PTRE)~\cite{cklee2015jcp,cwang2015sp,dzxu2016fp}
combined with full counting statistics (FCS)~\cite{llevitov1992jetp,llevitov1996jmp,mesposito2009rmp}, detailed at Sec. II.
The NE-PTRE can be reduced to the Redfield and noninteracting blip approximation (NIBA) schemes with weak and strong system-bath couplings, respectively.
At Sec. III A,  the NDTC is analyzed based on the two-terminal setup, and the mechanism of the NDTC is proposed in strong qubit-bath coupling limit.
Moreover, the partially strong qubit-bath interaction is sufficient to exhibit the NDTC.
At Sec. III B, the heat amplification is analyzed within the three-terminal setup,
and {the strong interaction of either the left or right qubit with the corresponding thermal bath can exhibit the dramatic heat amplification effect.}
Finally, we give a summary at Sec. IV.

\section{Model and method}
In this section, we first describe the nonequilibrium two-qubits model.
Then, by applying the polaron transformation we derive the nonequilibrium polaron-transformed Redfield equation to
analyze the dynamics of qubits density matrix.
Next, we obtain the expression of steady state heat flux by combining the NE-PTRE and full counting statistics.
Under the thermodynamic bias of thermal baths, the steady state heat transfer can be exploited.

%%==========================================
%%==========================================

\subsection{Nonequilibrium two-qubits model}
The nonequilibrium two-qubits system consists of two coupled two-level-system (TLS), each interacting with an individual thermal bath in Fig.~\ref{fig1}(a).
The two-qubits model is described as
\begin{eqnarray}~\label{ham0}
\hat{H}=U\hat{\sigma}^L_z\hat{\sigma}^R_z+\sum_{v=L,R}\hat{H}_v,
\end{eqnarray}
where $U$ is the inter-qubit coupling strength, the Pauli operators are expressed as
$\hat{\sigma}^v_z=|\uparrow_v{\rangle}{\langle}\uparrow_v|-|\downarrow_v{\rangle}{\langle}\downarrow_v|$
and
$\hat{\sigma}^v_x=|\uparrow_v{\rangle}{\langle}\downarrow_v|+|\downarrow_v{\rangle}{\langle}\uparrow_v|$,
with $|\uparrow_v{\rangle}~(|\downarrow_v{\rangle})$ the excited (ground) state of the $v$th qubit.
$\hat{H}_v$ is the seminal spin-boson model
\begin{eqnarray}~\label{hv}
\hat{H}_v=\hat{H}^v_b+\frac{\varepsilon_v}{2}\hat{\sigma}^v_z+\frac{\Delta_v}{2}\hat{\sigma}^v_x+
\sum_k(g_{kv}\hat{b}^{\dag}_{kv}+g^*_{kv}\hat{b}_{kv})\hat{\sigma}^v_z,
\end{eqnarray}
where $\hat{H}^v_b=\sum_{k,v}\omega_k\hat{b}^{\dag}_{kv}\hat{b}_{kv}$ stands for the $v$th noninteracting bosonic bath,
with $\hat{b}^{\dag}_{kv}~(\hat{b}_{kv})$ creating (annihilating) one phonon with energy $\omega_k$,
$\varepsilon_v$ is the qubit splitting energy, $\Delta_v$ is the tunneling strength
and $g_{kv}$ is the coupling strength between the $v$th qubit and the corresponding bath.
{The $v$th thermal bath is typically characterized by the spectral function $J_v(\omega)=4\pi\sum_k|g_{k,v}|^2\delta(\omega-\omega_k)$.
Here, we specify the spectral function as super-Ohmic form
$J_v(\omega)=\pi\alpha_v{\omega^3}e^{-\omega/\omega_c}/{\omega^2_c}$,
with $\alpha_v$ the qubit-bath coupling strength  and $\omega_c$ the cutoff frequency of the bath,
which has been extensively included in the quantum energy transfer~\cite{dzxu2016fp,anazir2009,sjang2013njp}.}
It is well-known that the spin-boson model was initially proposed to analyze the dissipative dynamics of quantum systems~\cite{agleggett1987rmp},
%And it was later extended to the quantum optics~\cite{hjCarmichael1999book,ybliu2017np},
%quantum phase transition~\cite{rbulla2008rmp,yyzhang2010,achin2011prl,zcai2014prl}
%and quantum measurement~\cite{hzheng2008prl,qai2010pra,she2017pra}.
and later extended to the nonequilibrium regime to
analyze the quantum thermal transfer~\cite{dsegal2005prl,dsegal2006prb,dsegal2008prl,dsegal2011prb,ksaito2013prl,dsegal2014pre,dsegal2015prl,akato2015jcp,jliu2017pre,zqiang2017prb}.

From Eq.~(\ref{hv}), it is known that qubits linearly interact with thermal bath.
Hence, to capture the multi-phonons effect in thermal transfer processes,
we apply a canonical transformation ("polaron transformation") to obtain the modified Hamiltonian
$\hat{H}^{\prime}=\hat{U}^{\dag}\hat{H}\hat{U}$~\cite{rsilbey1984jcp,rharris1985jcp,hzheng2004epjb}, where
the unitary operator is $\hat{U}=\exp{(i\sum_v\hat{B}_v\hat{\sigma}^v_z/2)}$ ,
with the collective phononic momentum
$\hat{B}_v=2i\sum_k(g_{kv}/\omega_k\hat{b}^{\dag}_{kv}-g^*_{kv}/\omega_k\hat{b}_{kv})$.
Then, the modified system Hamiltonian is given by
$\hat{H}^{\prime}=\hat{H}^{\prime}_s+\sum_{v}(\hat{H}^v_b+\hat{V}^{\prime}_v)$,
where the transformed qubits Hamiltonian is
\begin{eqnarray}~\label{hsp}
\hat{H}^{\prime}_s=U\hat{\sigma}^L_z\hat{\sigma}^R_z+\sum_{v=L,R}(\frac{\varepsilon_v}{2}\hat{\sigma}^v_z+\frac{\eta_v\Delta_v}{2}\hat{\sigma}^v_x),
\end{eqnarray}
where the renormalization factor $\eta_v={\langle}\cos\hat{B}_v{\rangle}$ is specified as
$\eta_v=\exp[-2\sum_k|g_{kv}/\omega_k|^2(2n_k+1)]$.
Moreover, the modified qubit-bath interaction is given by
\begin{eqnarray}~\label{vv}
\hat{V}^\prime_v=\frac{\Delta_v}{2}[\hat{\sigma}^v_x(\cos\hat{B_v}-\eta_v)+\hat{\sigma}^v_y\sin\hat{B}_v].
\end{eqnarray}
By expanding the interaction terms at Eq.~(\ref{vv}) as
$\hat{\sigma}^v_x(\cos\hat{B_v}-\eta_v)=\hat{\sigma}^v_x[(1-\eta_v)+\sum_{n=1}(\hat{B_v})^{2n}/(2n)!]$
and $\hat{\sigma}^v_y\sin\hat{B}_v=\hat{\sigma}^v_y[\sum_{n=0}(\hat{B_v})^{2n+1}/(2n+1)!]$,
it is easily found that multiple phonon effect is now explicitly included in heat transfer processes.

\subsection{Nonequilibrium polaron-transformed Redfield equation}
We apply the nonequilibrium polaron-transformed Redfield equation, one type of quantum master equation,
to study the dynamics of qubits density matrix.
Traditionally, the qubit-bath interaction at Eq.~(\ref{hv}) is directly perturbed to obtain the Redfield equation,
which is proper for weak qubit-bath coupling~\cite{dsegal2005prl,dsegal2006prb}.
While for NE-PTRE, it has been intensively applied in quantum heat transfer and heat engines~\cite{cklee2015jcp,cwang2015sp,dzxu2016fp}.
By perturbing the modified qubit-bath interaction at Eq.~(\ref{vv}), we are able to analyze dynamical behaviors
in the broad coupling region.
Specifically, based on the Born approximation, the total density matrix in the polaron framework is decomposed as
$\hat{\rho}_{tot}(t){\approx}\hat{\rho}_s(t){\otimes}\hat{\rho_b}$, where $\hat{\rho}_s(t)$ is the reduced density operator of qubits
and $\hat{\rho_b}=\exp{(-\sum_v\hat{H}^v_b/k_BT_v)}/\textrm{Tr}_b\{e^{-\sum_v\hat{H}^v_b/k_BT_v}\}$ is the equilibrium distribution operator of thermal baths.
Moreover, by considering the Markovian approximation, the NE-PTRE is obtained as
\begin{eqnarray}~\label{ptre1}
\frac{d\hat{\rho}_s(t)}{{d}t}&=&-i[\hat{H}^{\prime}_s,\hat{\rho}_s(t)]+\sum_{v=L,R;a=x,y;\omega,\omega^{\prime}}\\
&&{\times}(\Gamma^v_{a,+}(\omega)[\hat{P}^v_a(\omega)\hat{\rho}_s(t),\hat{P}^v_a(\omega^{\prime})]+H.c.),\nonumber
\end{eqnarray}
where the projecting operators are $\hat{\sigma}^v_a(-\tau)=\sum_{\omega}\hat{P}^v_a(\omega)e^{i\omega\tau}$,
transition rates are
\begin{eqnarray}~\label{gv}
\Gamma^v_{a,\pm}(\omega)=\int^\infty_0d{\tau}C^v_a(\pm\tau)e^{i\omega\tau}.
\end{eqnarray}
and the correlation functions are given by
\begin{eqnarray}
C^v_x(\tau)&=&(\frac{\eta_v\Delta_v}{2})^2[\cosh{Q_v(\tau)}-1]\\
C^v_y(\tau)&=&(\frac{\eta_v\Delta_v}{2})^2\sinh{Q_v(\tau)},\nonumber
\end{eqnarray}
%\begin{eqnarray}
%C^v_x(\tau)&=&(\frac{\Delta_v}{2})^2[{\langle}\cos\hat{B}_v(\tau)\cos\hat{B}_v{\rangle}-(\eta_v)^2]\\
%&=&(\frac{\eta_v\Delta_v}{2})^2[(e^{Q_v(\tau)}+e^{-Q_v(\tau)})/2-1]\\
%C^v_y(\tau)&=&(\frac{\Delta_v}{2})^2{\langle}\sin\hat{B}_v(\tau)\sin\hat{B}_v{\rangle}\nonumber\\
%&=&(\frac{\eta_v\Delta_v}{2})^2(e^{Q_v(\tau)}-e^{-Q_v(\tau)})/2,\nonumber
%\end{eqnarray}
with the propagating phase of phonons
%\begin{eqnarray}
%Q^v_{\pm}(\tau)=4\sum_{k}|\frac{g_{kv}}{\omega_k}|^2[(1{\mp}\cos\omega_k\tau)(2n_k+1){\pm}i\sin\omega_k\tau]
%\end{eqnarray}
\begin{eqnarray}
Q_v(\tau)=4\sum_{k}|\frac{g_{kv}}{\omega_k}|^2[\cos\omega_k\tau(2n_k+1)-i\sin\omega_k\tau],
\end{eqnarray}
and the Bose-Einstein distribution function $n_{kv}=1/[\exp{(\omega_k/k_BT_v)}-1]$.
%\begin{eqnarray}
%\frac{d\hat{\rho}_s(t)}{{d}t}&=&-i[\hat{H}^{\prime}_s,\hat{\rho}_s(t)]\\
%&&+\sum_{v=L,R;a=x,y}\int^\infty_0d\tau(C^v_a(\tau)[\hat{\sigma}^v_a(-\tau)\hat{\rho}_s(t),\hat{\sigma}^v_a]+H.c.),\nonumber
%\end{eqnarray}
%where the correlation functions are given by
%\begin{eqnarray}
%C^x_v(\tau)&=&(\frac{\Delta_v}{2})^2[{\langle}\cos\hat{B}_v(\tau)\cos\hat{B}_v{\rangle}-(\eta_v)^2]\\
%&=&(\frac{\Delta_v}{2})^2[(e^{-Q^v_+(\tau)}+e^{-Q^v_-(\tau)})/2-(\eta_v)^2]\\
%C^y_v(\tau)&=&(\frac{\Delta_v}{2})^2{\langle}\sin\hat{B}_v(\tau)\sin\hat{B}_v{\rangle}\nonumber\\
%&=&(\frac{\Delta_v}{2})^2(e^{-Q^v_+(\tau)}-e^{-Q^v_-(\tau)}),\nonumber
%\end{eqnarray}
{The transition rate $\Gamma^v_{y,\pm}(\omega)$ describes that odd phonons are involved in the heat exchange processes}
between the $v$th qubit and the corresponding bath with transition energy $\omega$ at Eq.~(\ref{gv}).
For the lowest order of correlation function $C^v_y(\pm\tau)$ as $(\frac{\eta_v\Delta_v}{2})^2Q_v(\pm\tau)$,
the rates are reduced to
$\textrm{Re}[\Gamma^v_{y,-}(\omega)]=(\frac{\eta_v\Delta_v}{2\omega})^2J_v(\omega)n_v(\omega)$
and
$\textrm{Re}[\Gamma^v_{y,+}(\omega)]=(\frac{\eta_v\Delta_v}{2\omega})^2J_v(\omega)(1+n_v(\omega))$,
with the spectral function of the $v$th bath as $J_v(\omega)=4\pi\sum_k|g_{k,v}|^2\delta(\omega-\omega_k)$
and $n_v(\omega)=1/[e^{\omega/k_BT_v}-1]$,
which exhibits sequential excitation and relaxation of the $v$th qubit by absorbing and emitting one phonon, accordingly.
While the transition rate $\Gamma^v_{x,\pm}(\omega)$ shows even phonons involved transfer processes.
The lowest order of correlation function $C^v_x(\pm\tau)$ is given by $(\frac{\eta_v\Delta_v}{2})^2Q^2_v(\pm\tau)$.
Hence, the reduced rates are obtained as
$\textrm{Re}[\Gamma^v_{x,\pm}(\omega)]
=(\frac{\eta_v\Delta_v}{2})^2\int^\infty_0\frac{d{\omega_1}}{4\pi}Q_v(\pm\omega_1)Q_v(\pm\omega\mp\omega_1)$,
with $Q_v(\omega_1)=\int{d\tau}e^{i\omega_1\tau}Q_v(\tau)$.
For $\textrm{Re}[\Gamma^v_{x,\pm}(\omega)]$, as the $v$th qubit is excited(relaxed) with energy gap $\omega$,
it simultaneously absorbs(emits) two phonons with energy $\omega_1$ and $\omega-\omega_1$ from(to) the $v$th bath.

%Describe the physical process through the systematic expansion of correlation function.
In previous works of nonequilibrium spin-boson model, the NE-PTRE was found to unify steady state behaviors
in the weak and strong qubit-bath coupling limit, respectively~\cite{cwang2015sp,dzxu2016fp,cwang2017pra}.
Here, we note that in the two-qubits system the NE-PTRE can also find such correspondence.
Specifically, in the weak coupling regime ($|g_{k,v}/\omega_k|^2{\approx}0$), the renormalization factor becomes $\eta_v{\approx}1$.
Hence, the qubits Hamiltonian is simplified as
$\hat{H}^{\prime}_s=U\hat{\sigma}^1_z\hat{\sigma}^2_z+\sum_v(\frac{\varepsilon_v}{2}\hat{\sigma}^v_z+\frac{\Delta_v}{2}\hat{\sigma}^v_x)$.
And the lowest order of transformed qubit-bath interaction $\hat{V}^\prime_v{\approx}\hat{\sigma}^v_y\hat{B}_v$ dominates the dynamics,
which results in the correlation functions
$C^v_x(\tau){\approx}0$ and $C^v_y(\tau){\approx}(\frac{\Delta_v}{2})^2Q_v(\tau)$.
Under the basis $\{|\phi_n{\rangle}\}$ with $\hat{H}^\prime_s|\phi_n{\rangle}=E_n|\phi_n{\rangle}$,
the dynamical equation at Eq.~(\ref{ptre1}) after long-time evolution can be reduced to
\begin{eqnarray}~\label{redfield1}
\frac{d\rho_{nn}}{dt}&=&\sum_{m{\neq}n,v}\frac{J_v(E_{nm})}{2}n_v(E_{nm})(\frac{\Delta_v}{E_{nm}})^2\sigma^{nm}_{y,v}\sigma^{mn}_{y,v}\rho_{mm}\nonumber\\
&&-\sum_{m{\neq}n,v}\frac{J_v(E_{mn})}{2}n_v(E_{mn})(\frac{\Delta_v}{E_{nm}})^2\sigma^{nm}_{y,v}\sigma^{mn}_{y,v}\rho_{nn},\nonumber\\
\end{eqnarray}
with the density matrix element $\rho_{nn}={\langle}\phi_n|\hat{\rho}_s|\phi_n{\rangle}$,
the energy gap $E_{nm}=E_n-E_m$, $n_v(E_{nm})=1/[\exp{(E_{nm}/k_BT_v)}-1]$ and $\sigma^{nm}_{y,v}={\langle}\phi_n|\hat{\sigma}^v_y|\phi_m{\rangle}$.
Moreover, from the commutating relationship
${\langle}\phi_n|[\hat{\sigma}^v_z,\hat{H}^\prime_s]|\phi_m{\rangle}=i\Delta_v\sigma^{nm}_{y,v}=E_{nm}\sigma^{nm}_{z,v}$,
we conclude that Eq.~(\ref{redfield1}) is equivalent with the Redfield equation at Eq.~(\ref{app-a-redfield}) (see appendix A for the detail).

While in the strong qubit-bath interaction regime ($|g_{k,v}/\omega_k|^2{\gg}1$), the renormalization factor becomes $\eta_v{\approx}0$.
Hence the qubits Hamiltonian is simplified as $\hat{H}^{\prime}_s=U\hat{\sigma}^1_z\hat{\sigma}^2_z+\sum_v\frac{\varepsilon_v}{2}\hat{\sigma}^v_z$,
and the qubit-bath interaction is give by $\hat{V}^{\prime}_v=\frac{\Delta_v}{2}[\hat{\sigma}^v_x\cos{B}_v+\hat{\sigma}^v_y\sin{\hat{B}_v}]$.
The correlation functions are reduced to $C^v_x(\tau)=C^v_y(\tau)=\Delta^2_v\eta^2_ve^{Q_v(\tau)}/8$.
It should be noted that though $\eta_v{\approx}0$, $\eta^2_ve^{Q_v(\tau)}$ will be kept as a finite value.
Consequently, NE-PTRE at Eq.~(\ref{redfield1}) is reduced to the nonequilibrium NIBA limit at Eq.~(\ref{app-b-NIBA}) (see appendix B for the detail).

%describe the Redfield and NIBA limit.
%with
%\begin{eqnarray}
%{\langle}e^{{\pm}i\hat{B}^{-\chi}_v(\tau)}e^{{\mp}i\hat{B}^{\chi}_v}{\rangle}
%=\exp(-4\sum_{k}|\frac{g_{kv}}{\omega_k}|^2[(1-\cos\omega_k(\tau-\chi))(2n_k+1)+i\sin\omega_k(\tau-\chi)])
%\end{eqnarray}

%%==========================================
\begin{figure}[tbp]
\begin{center}
%\centering
%\vspace{-2.2cm}
\includegraphics[scale=0.5]{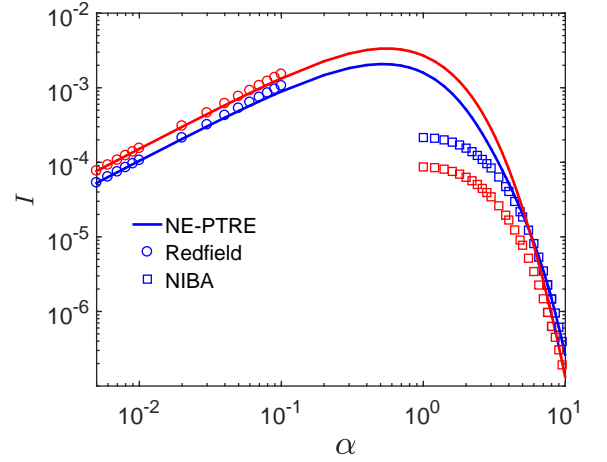}
\vspace{-1.2cm}
\end{center}
\caption{(Color online) Comparison of steady state heat currents by tuning qubit-bath coupling strength $\alpha$
from the NE-PTRE (blue-dark and red-light solid lines), Redfield equation (blue-dark and red-light circles)
and nonequilibrium NIBA scheme (blue-dark and red-light squares), respectively.
The energy bias is selected as $\varepsilon=0$ for the blue-dark solid line, circles and squares,
and $\varepsilon=1$ for the red-light solid line, circles and squares.
The other parameters are given by $\Delta=1$, $U=0.1$, $\omega_c=5$, $T_L=1.5$ and $T_R=0.5$.
}~\label{fig2}
\end{figure}
%%==========================================
\subsection{Steady state heat current}
We combine NE-PTRE with FCS to count the heat flow into thermal baths~\cite{mesposito2009rmp,cwang2015sp}.
FCS is considered as a powerful method to investigate the full information of current fluctuations,
which was initially proposed to analyze the charge fluctuations~\cite{llevitov1992jetp,llevitov1996jmp}.
For the nonequilibrium spin-boson system, FCS has been extensively applied to study quantum heat transport at steady state or driven by the geometric phase~\cite{jren2010prl,jren2013prb,cuchiyama2014pre,cwang2017pra,jliu2018jcp}.
Generally, by employing a two-time measurement protocol during the time interval $[0,t]$, the generating function to characterize heat transfer
is expressed as~\cite{hmfriedman2018njp}
$\mathcal{Z}(\chi_v,t)=\textrm{Tr}\{e^{i\chi_v\hat{H}^v_b(0)}e^{-i\chi_v\hat{H}^v_b(t)}\hat{\rho}_{tot}(0)\}$,
where $\chi_v$ is the counting field parameter to count energy into the $v$th bath, $\hat{\rho}_{tot}(0)$ is the initial density operator of the whole system,
and the operator $\hat{A}(t)$ in the Heisenberg representation is $\hat{A}(t)=e^{i\hat{H}t}\hat{A}e^{-i\hat{H}t}$.
Moreover, by assuming $[\hat{\rho}_{tot}(0),\hat{H}^v_b]=0$, the generating function can be re-expressed as
\begin{eqnarray}~\label{zchi}
\mathcal{Z}(\chi_v,t)&=&\textrm{Tr}\{\hat{M}_{-\chi_v}(t)\hat{\rho}_{tot}(0)\hat{M}^{\dag}_{\chi_v}(t)\}\\
&=&\textrm{Tr}\{\hat{\rho}_{\chi_v}(t)\},\nonumber
\end{eqnarray}
where the generalized propagators are $\hat{M}_{-\chi_v}(t)=e^{-i\hat{H}_{-\chi_v}t}$
and $\hat{M}^{\dag}_{\chi_v}(t)=e^{i\hat{H}_{\chi_v}t}$,
with $\hat{H}_{\chi_v}=e^{\frac{i\chi_v}{2}\hat{H}^v_b}\hat{H}e^{-\frac{i\chi_v}{2}\hat{H}^v_b}$.
Then, the cumulant generating function at steady state can be obtained as
\begin{eqnarray}~\label{gchi}
G(\chi_v)=\lim_{t{\rightarrow}\infty}\frac{1}{t}\textrm{ln}\mathcal{Z}(\chi_v,t).
\end{eqnarray}
Consequently, the $n$th-order current cumulant is given by
\begin{eqnarray}~\label{in}
I^{(n)}=\frac{{\partial}^nG(\chi_v)}{{\partial}(i\chi_v)^n}\Big{|}_{\chi_v=0}.
\end{eqnarray}
In particular, the lowest-order cumulant is the heat current
\begin{eqnarray}
I=\frac{{\partial}G(\chi_v)}{{\partial}(i\chi_v)}\Big{|}_{\chi_v=0}.
\end{eqnarray}
Therefore, once we obtain the dynamics of the counting-field dependent density matrix of the qubits system, we are able to obtain full information of current fluctuations.

%describe the full counting statistics and applications. convenience in heat transfer.
Then, we apply FCS to study the heat transfer in two-qubit system.
The counting-field dependent Hamiltonian $\hat{H}(\chi)=e^{\frac{i\chi}{2}\hat{H}_{b,R}}\hat{H}e^{-\frac{i\chi}{2}\hat{H}_{b,R}}$ is given by
$\hat{H}_{{\chi}}=U\hat{\sigma}^L_z\hat{\sigma}^R_z+\sum_{v=L,R}\hat{H}_v(\chi)$,
where the sub-Hamiltonian is
\begin{eqnarray}
\hat{H}_v(\chi)&=&\hat{H}^v_b+\frac{\varepsilon_v}{2}\hat{\sigma}^v_z+\frac{\Delta_v}{2}\hat{\sigma}^v_x\\
&&+\hat{\sigma}^v_z\sum_k(g_{kv}e^{\frac{i\chi}{2}\omega_k\delta_{v,R}}\hat{b}^{\dag}_{kv}+H.c.),\nonumber
\end{eqnarray}
with $\hat{H}^v_b=\sum_k\omega_k\hat{b}^{\dag}_{k,v}\hat{b}_{k,v}$.
Furthermore, we apply the modified canonical operator
$\hat{U}_{\chi}=\exp[\frac{i}{2}(\hat{B}_L\hat{\sigma}^L_z+\hat{B}_R(\chi)\hat{\sigma}^R_z)]$ to $\hat{H}_\chi$
as $\hat{H}^{\prime}_{\chi}=\hat{U}^{\dag}_{\chi}\hat{H}_{\chi}\hat{U}_{\chi}$,
resulting in
\begin{eqnarray}~\label{fcs-hv}
\hat{H}^{\prime}_{\chi}=\hat{H}^{\prime}_s+\sum_{v=L,R}(\hat{H}^v_b+\hat{V}_v({\chi})),
\end{eqnarray}
with
\begin{eqnarray}
\hat{V}_v({\chi})=\frac{\Delta_v}{2}[\hat{\sigma}^v_x(\cos\hat{B^{\chi}_v}-\eta_v)+\hat{\sigma}^v_y\sin\hat{B}^{\chi}_v],
\end{eqnarray}
and
$\hat{B}^{\chi}_v=2i\sum_k(\frac{g_{kv}}{\omega_k}e^{\frac{i\chi}{2}\omega_k\delta_{v,R}}\hat{b}^{\dag}_{kv}-H.c.)$.
Following a similar procedure, the total density matrix is decomposed as
$\hat{\rho}^\chi_{tot}(t)=\hat{\rho}_{\chi}(t){\otimes}\hat{\rho}_b$.
The counting-field dependent NE-PTRE is obtained as
\begin{eqnarray}~\label{ptre-chi}
\frac{d\hat{\rho}_{\chi}(t)}{dt}&=&-i[\hat{H}^{\prime}_s,\hat{\rho}_{\chi}(t)]+\sum_{v=L,R;a=x,y;\omega,\omega^{\prime}}\\
&&{\times}[(\Gamma^v_{a,-}(\chi,\omega)+\Gamma^v_{a,+}(\chi,\omega^{\prime}))\hat{P}^v_a(\omega^{\prime})\hat{\rho}_{\chi}(t)\hat{P}^v_a(\omega)\nonumber\\
&&-(\Gamma^v_{a,+}(\omega)\hat{P}^v_a(\omega^{\prime})\hat{P}^v_a(\omega)\hat{\rho}_{\chi}(t)+H.c.)],\nonumber
\end{eqnarray}
where the modified transition rates are
\begin{eqnarray}~\label{twoterminal-rate}
\Gamma^v_{a,\pm}(\chi,\omega)=\int^{\infty}_0d{\tau}C^v_a(\chi,\pm\tau)e^{i\omega\tau}.
\end{eqnarray}
and the correlation functions are $C^v_a(\chi,\pm\tau)=C^v_a(\pm\tau-\chi\delta_{v,R})$.
As the counting field parameter vanishes, the dynamical equation at Eq.~(\ref{ptre-chi}) is reduced to the standard version
at Eq.~(\ref{ptre1}).

Now, based on the definition of generating function at Eq.~(\ref{zchi}) and current fluctuations at Eq.~(\ref{in}), it is ready to analyze the steady state heat transfer by applying the dynamics of qubits density matrix at Eq.~(\ref{ptre-chi}).
In the following, we simplify system parameters as $\Delta_v=\Delta$, $\varepsilon_v=\varepsilon$.
Here, the first question we should answer is: \emph{does the heat current derived from the counting-field dependent NE-PTRE unifies
the results based on the Redfield and nonequilibrium NIBA schemes?}
In Fig.~\ref{fig2}, it clearly demonstrates that both at resonance ($\varepsilon=0$) and off-resonance (e.g., $\varepsilon=1$),
heat currents perfectly bridge the weak and strong coupling results.
Therefore, it is safe to apply the NE-PTRE in a wide qubit-bath coupling regime to analyze nonequilibrium steady state behaviors.

%explain the unified result for heat transfer, and prove our method can unify the heat current. Then we apply to the NDTC with weak and strong coupling
%and later transistor.

%At resonance ($\varepsilon_v=0$), the analytical expression of heat current in strong coupling regime is given by
%The heat current is given by
%\begin{eqnarray}
%I&=&2U\frac{\kappa_R(2U)\kappa_L(-2U)-\kappa_R(-2U)\kappa_L(2U)}{\kappa_L(2U)+\kappa_R(2U)+\kappa_L(-2U)+\kappa_R(-2U)}
%\end{eqnarray}

%%==========================================
\begin{figure}[tbp]
\begin{center}
%\vspace{-1.0cm}
\includegraphics[scale=0.42]{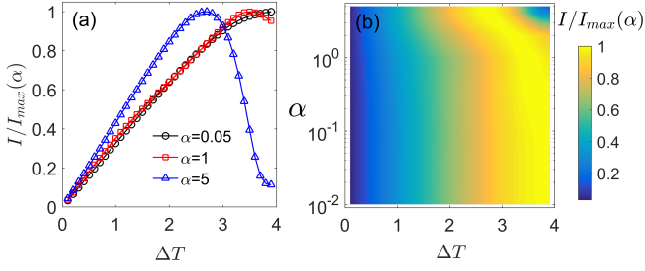}
\vspace{-2.0cm}
\end{center}
\caption{(Color online) Normalized heat current $I/I_{max}(\alpha)$ by
(a) tuning temperature difference ${\Delta}T=T_L-T_R$ with typical qubit-bath coupling strength $\alpha$
and (b) both modulating ${\Delta}T$ and $\alpha$ in a 3D view,
with $I_{\max}(\alpha)=\max_{{\Delta}T}\{I\}$ for a given $\alpha$.
The temperatures are given by $T_L=T_0+{\Delta}T/2$ and $T_R=T_0-{\Delta}T/2$ with $T_0=2$.
The other parameters are  $\varepsilon=1$, $\Delta=1$, $U=0.1$, $\omega_c=5$,
}~\label{fig3}
\end{figure}
%%==========================================

%%==========================================
\begin{figure}[tbp]
%\begin{center}
\hspace{-4.0cm}
\includegraphics[scale=0.18]{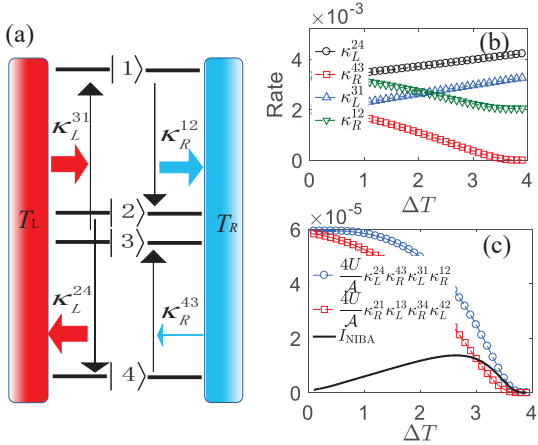}
%\vspace{-1.0cm}
%\end{center}
\caption{(Color online) (a) {Schematic illustration of the heat transfer process for one current component
$4U\kappa^{24}_L\kappa^{43}_R\kappa^{31}_L\kappa^{12}_R/\mathcal{A}$ at Eq.~(\ref{current-2})} in strong qubit-bath coupling limit
with the states $|1{\rangle}=|\uparrow\uparrow{\rangle}$, $|2{\rangle}=|\uparrow\downarrow{\rangle}$,
$|3{\rangle}=|\downarrow\uparrow{\rangle}$, $|4{\rangle}=|\downarrow\downarrow{\rangle}$, the transition rate $\kappa^{ij}_{v}~(v=L,R)$ at Eq.~(\ref{ratenm}),
{and horizontal (red-light and blue-dark) arrows with wide and thin sizes representing different amplitudes of $\kappa^{ij}_{v}$};
(b) {the corresponding transition rates}  and (c) two loop currents and $I_{\mathrm{NIBA}}$ at Eq.~(\ref{current-2}) as a function of ${\Delta}T$ with $\alpha=5$.
The other parameters are the same as in Fig.~\ref{fig3}.
}~\label{fig4}
\end{figure}
%%==========================================

\section{Results and discussions}

\subsection{Negative differential thermal conductance}
Negative differential thermal conductance (NDTC), as analogue of the negative differential conductance in electronics,
traditionally describes the process that heat current is suppressed by increasing the temperature bias in a two-terminal setup~\cite{bli2006prl}.
It has been extensively applied to investigate quantum thermal transport in phononics~\cite{nbli2012rmp}.
However, the NDTC is less exploited for nonequilibrium two-qubits systems.

Here, we first analyze the NDTC in a biased two-qubits model by tuning qubit-bath coupling within the NE-PTRE in Fig.~\ref{fig3}(a).
For the weak coupling case ($e.g., \alpha=0.05$), It is shown that the normalized heat current $I/I_{max}(\alpha)$ becomes monotonically enhanced
by tuning the temperature bias $\Delta{T}$.
As the coupling strength is enlarged, the NDTC appears and finally becomes significant ($e.g., \alpha=5$).
Hence, we conclude that the NDTC exists in the strongly-coupled two-spin-boson model.
To give a comprehensive picture of the NDTC under the modulation of the temperature bias and the coupling strength, we plot Fig.~\ref{fig3}(b).
It is found that at the strong coupling regime ($\alpha{\gtrsim}0.5$) the $I/I_{max}(\alpha)$ shows the turnover behavior, corresponding to the NDTC.
%While as $\alpha{<}0.5$ the $I/I_{max}(\alpha)$ exhibits monotonic increase.

{In previous works with strong qubit-bath coupling, it is known that the single qubit is unable to exhibit the NDTC~\cite{cwang2015sp}.
While for the three-qubits case, the NDTC is clearly shown~\cite{cwang2018pra}.
Therefore, considering the result obtained in the present paper with the two-qubits system,
we conclude that multi-qubits and strong qubit-bath interaction are necessary conditions to generate the NDTC.
It should be noted that the current conclusion is limited to such (qubit-qubit and qubit-bath) interactions of the Hamiltonian at Eq.~(\ref{ham0})
and Eqs.~(1-2) in Ref.~\cite{cwang2018pra}.
While for {other kinds of quantum spin systems}, the NDTC can also be unravelled with weak qubit-bath coupling, e.g., quantum thermal transistor~\cite{kjoulain2016prl}
and spin seebeck transistor~\cite{jren2013prb,jren2013prb2,jren2013prb3}.}

\subsubsection{The mechanism of NDTC}
Next, we try to provide some insight into the underlying mechanism of the NDTC in strong qubit-bath coupling limit.
It is known that with strong coupling the nonequilbrium NIBA can at least qualitatively describe heat transfer processes.
{Under the condition $\varepsilon_v=\varepsilon$, the steady state heat current can be approximately expressed as (see Eq.~(\ref{I_strong_1}) at appendix B)
\begin{eqnarray}~\label{current-2}
I_{\mathrm{NIBA}}&=&\frac{4U}{\mathcal{A}}(\kappa^{24}_L\kappa^{43}_R\kappa^{31}_L\kappa^{12}_R-\kappa^{21}_R\kappa^{13}_L\kappa^{34}_R\kappa^{42}_L),
\end{eqnarray}
where the coefficient $\mathcal{A}$ is
\begin{eqnarray}
\mathcal{A}&=&[\kappa^{12}_L\kappa^{42}_L\kappa^{21}_R+\kappa^{21}_L\kappa^{24}_L\kappa^{42}_R+\kappa^{21}_L\kappa^{42}_L\kappa^{21}_R
+\kappa^{21}_L\kappa^{42}_L\kappa^{12}_R\nonumber\\
&&+\kappa^{21}_L\kappa^{24}_L\kappa^{12}_R+(\kappa^{12}_L+\kappa^{12}_R)(\kappa^{24}_L\kappa^{42}_R+\kappa^{24}_L\kappa^{24}_R/2)]\nonumber\\
&&+[L{\leftrightarrow}R],\nonumber
\end{eqnarray}
which is invariant by exchanging the subindex $L$ with $R$.
$\kappa^{2i}_v=\kappa^{3i}_v$, $\kappa^{i2}_v=\kappa^{i3}_v$ and
the transition rate is given by
\begin{eqnarray}~\label{ratenm}
\kappa^{ij}_v=(\frac{\eta_v\Delta_v}{2})^2\int^\infty_{-\infty}d{\tau}e^{Q_v(\tau)}e^{iE_{ij}\tau}
\end{eqnarray}
with the energy gap is $E_{ij}=E_i-E_j$ with
$(U\hat{\sigma}^L_z\hat{\sigma}^R_z+\sum_v\frac{\varepsilon_v}{2}\hat{\sigma}^v_z)|i{\rangle}=E_i|i{\rangle}~(i=1,2,3,4)$.
From Eq.~(\ref{current-2}), it is clearly shown that the heat current is contributed by two loop currents
${4U}\kappa^{24}_L\kappa^{43}_R\kappa^{31}_L\kappa^{12}_R/{\mathcal{A}}$
and ${4U}\kappa^{21}_R\kappa^{13}_L\kappa^{34}_R\kappa^{42}_L/{\mathcal{A}}$.

Then, we analyze the heat current in two limiting temperature bias regimes to exploit the NDTC.
For the low bias ${\Delta}T{\ll}T_0$, the heat current is approximately expressed as
\begin{eqnarray}
I_{\mathrm{NIBA}}{\approx}\frac{(4U)^2{\Delta}T}{\mathcal{A}T^2_0}\kappa^{24}_L\kappa^{43}_R\kappa^{31}_L\kappa^{12}_R.
\end{eqnarray}
Moreover, the term $\kappa^{24}_L\kappa^{43}_R\kappa^{31}_L\kappa^{12}_R/{\mathcal{A}}$ is nearly constant,
shown with { blue-dark dashed line} with circles in Fig.~\ref{fig4}(c).
Hence, $I_{\mathrm{NIBA}}$ is proportional to ${\Delta}T$, which is exhibited with  black solid line}.
While for the large temperature bias ${\Delta}T{\approx}2T_0$, i.e. $T_L{\approx}2T_0$ and $T_R{\approx}0$,
The transition from $|\downarrow\downarrow{\rangle}$ to $|\downarrow\uparrow{\rangle}$ assisted by the right bath is significantly suppressed($\kappa^{43}_R{\approx}0$),
which is schematically demonstrated in Fig.~\ref{fig4}(a) and also exhibited in Fig.~\ref{fig4}(b).
It is mainly attributed to the fact that phonons are difficult to excite in low temperature limit($T_R{\approx}0$).
Hence, the loop current ${4U}\kappa^{24}_L\kappa^{43}_R\kappa^{31}_L\kappa^{12}_R/{\mathcal{A}}$ becomes negligible, shown in Fig.~\ref{fig4}(c).
Similarly, the loop current
$\frac{4U}{\mathcal{A}}\kappa^{21}_R\kappa^{13}_L\kappa^{34}_R\kappa^{42}_L$ is also blocked,
due to the monotonic suppression of the transition rate $\kappa^{21}_{R}$(not shown here).
Therefore, we conclude that the heat current is strongly suppressed by the large temperature bias.
Combining the linear increase of the current in the linear response regime and dramatic suppression at large temperature bias,
the heat current should exhibit the turnover feature, which is the signature of the NDTC.
It should be noted that the analysis of the origin of the NDTC is approximate based on the nonequilibrium NIBA.

%More specifically, we study the steady state populations and transition rates in Fig.~\ref{fig4}(c) and Fig.~\ref{fig4}(d).
%By increasing $\Delta{T}$ in the low temperature bias regime, it is found that the change of $P_1$ is negligible and $P_2$ shows gradual enhancement.
%While for transition rates $\kappa^{12}_R$ and $\kappa^{21}_R$,
%they obey the detail balanced relationship as $\kappa^{21}_R=\exp{(-E_{12}/k_BT_R)}\kappa^{12}_R$.
%Then, by re-expressing the first component of heat current at Eq.~(\ref{current-1}) as
%$I_1=E_{12}\kappa^{12}_R(P_1-e^{-\frac{E_{12}}{k_BT_R}}P_2)$,
%it is found that though the rate $\kappa^{12}_R$ is suppressed, $(P_1-e^{-\frac{E_{12}}{k_BT_R}}P_2)$ compensates such loss,
%which finally results in the increase of $I_1$.
%In the large temperature bias ($e.g., {\Delta}T=4$), the vanishing values of the population $P_1$ and the factor $e^{-\frac{E_{12}}{k_BT_R}}$
%show no contribution to heat current.
%Hence, $I_1$ exhibits nonmonotonic feature in the wide regime of ${\Delta}T$.
%While for the second component of current at Eq.~(\ref{current-1}),
%it is expressed as $I_2=E_{34}\kappa^{43}_R(P_4-e^{\frac{E_{34}}{k_BT_R}}P_3)$.
%In the low bias of ${\Delta}T$, the rate $\kappa^{43}_R$ decreases monotonically,
%whereas $(P_4-e^{\frac{E_{34}}{k_BT_R}}P_3)$ is dramatically enhanced to finally increase the heat current.
%In the large temperature bias, both $\kappa^{43}_R$ and $P_3$ become negligible, which results in the vanish of $I_2$.
%Therefore, we provide an insight to explain the mechanism of the NDTC.

%%==========================================
\begin{figure}[tbp]
\begin{center}
%\vspace{-1.0cm}
\includegraphics[scale=0.4]{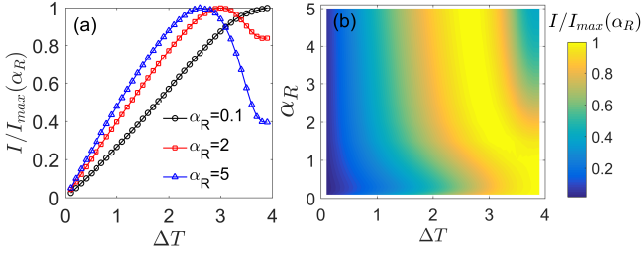}
%\vspace{-1.0cm}
\end{center}
\caption{(Color online) Normalized heat current $I/I_{max}(\alpha_R)$ (a) by tuning temperature bias ${\Delta}T$ with typical $\alpha_R$,
and (b) by modulating both ${\Delta}T$ and $\alpha_R$ in a 3D view,
with $I_{max}(\alpha_R)=\max_{{\Delta}T}\{I\}$ at given $\alpha_L$ and $\alpha_R$.
The left qubit-bath coupling strength is $\alpha_L=0.05$ and the other parameters are the same as in Fig.~\ref{fig3}.
}~\label{fig5}
\end{figure}
%%==========================================

\subsubsection{Partially strong qubit-bath interaction exhibits NDTC}
Next, we raise the question that whether the strong coupling between the $v$th qubit and the corresponding thermal bath
is sufficient to induce the negative differential thermal conductance?
By setting the weak left qubit-bath coupling (e.g., $\alpha_L=0.05$) in Fig.~\ref{fig5}(a),
we investigate the influence of the right qubit-bath interaction in the steady state heat current.
It is interesting to find that in strong coupling regime of $\alpha_R$, the NDTC also appears.
Thus, the strong coupling between the right qubit and bath is sufficient to exhibit the NDTC.
To give a comprehensive picture of the NDTC, we plot Fig.~\ref{fig5}(b).
It is found that as $\alpha_R{\gtrsim}1.2$, the NDTC can be observed by tuning the temperature bias.
{Hence, partially strong qubit-bath interaction is sufficient to exhibit the NDTC.}

%%==========================================
\begin{figure}[tbp]
\begin{center}
\includegraphics[scale=0.43]{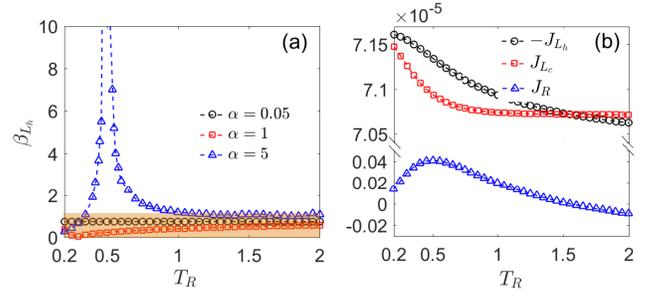}
\vspace{-3.0cm}
\end{center}
\caption{(Color online) (a) Heat amplification factor $\beta_{L_h}$ with typical qubit-bath couplings and (b) steady state currents with strong coupling ($\alpha=5$), by varying temperature of right bath $T_R$.
The other parameters are given by $\varepsilon=1$, $\Delta=1$, $U=0.1$, $\omega_c=5$, $T_{L_h}=2$ and $T_{L_c}=0.2$.
}~\label{fig6}
\end{figure}
%%==========================================
\subsection{Heat amplification}
Heat amplification, the key ingredient of quantum thermal transistor, shows a power to dramatically amplify heat flow via a tiny change of one terminal current
within a three-terminal setup~\cite{bli2006prl}.
It has been exploited in nonequilibrium three-qubits systems with both weak and strong qubit-bath coupling cases~\cite{kjoulain2016prl,cwang2018pra}.
In Fig.~\ref{fig1}(b), the amplification factor can be expressed as
\begin{eqnarray}~\label{beta-u}
\beta_{L_u}=|{\partial}I_{L_u}/{\partial}I_R|,u=h,c.
\end{eqnarray}
Considering the energy conservation $\sum_{v=L_h,L_c,R}I_v=0$,
the factor relationship with the $L_h$  and $L_c$ terminals can be established as
\begin{eqnarray}
\beta_{L_c}=|\beta_{L_h}+(-1)^{\theta}|,
\end{eqnarray}
where the phase $\theta=0$ as ${\partial}I_{L_u}/{\partial}I_R>0$
and $\theta=1$ as ${\partial}I_{L_u}/{\partial}I_R<0$.
Traditionally, the thermal transistor is able to work as $\beta_{L_u}>1$.
Here, we focus on the amplification effect in a two-qubits system, asymmetrically coupled to three thermal baths
(see the derivation of the dynamical equation  at appendix C).

We first investigate the influence of the qubit-bath interaction in the heat amplification factor  $\beta_{L_h}$ by tuning $T_R$ in Fig.~\ref{fig6}(a).
In the weak  coupling regime(e.g., $\alpha=0.05$), $\beta_{L_h}$ is below the unit, which implies the breakdown of heat amplification.
While in strong coupling regime (e.g., $\alpha=5$), it is interesting to observe a giant amplification factor in comparatively low temperature regime around $T_R{\approx}0.5$, and is dramatically suppressed as qubits is driven away from this critical temperature regime.
Moreover, in Fig.~\ref{fig6}(b) the heat current into the $R$th bath shows nonmonotonic behavior with strong coupling by modulating $T_R$.
From the definition of $\beta_{L_h}$ at Eq.~(\ref{beta-u}), it is understandable the amplification factor becomes divergent at the turnover point of $I_R$ (${\partial}I_R/{\partial}T_R=0$).
%The brief analysis of the turnover behavior of $I_R$ can be found in Fig.~\ref{append-c-fig1} at appendix C.
%The underlying mechanism of this turnover behavior of $I_R$ is extremely similar to the NDTC in two-terminal case
%(see Eq.~(\ref{app-c-IR}) and Fig.~\ref{append-c-fig1} at appendix C for the detail).
Therefore, we conclude that strong qubit-bath coupling is able to exhibit the significant heat amplification effect.

%%==========================================
\begin{figure}[tbp]
\begin{center}
\vspace{-1.0cm}
\includegraphics[scale=0.4]{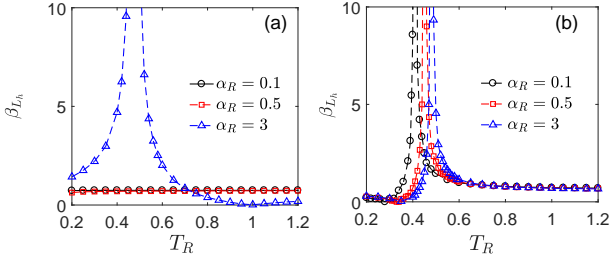}
\vspace{-2.0cm}
\end{center}
\caption{(Color online) Heat amplification factor $\beta_{L_h}$ as a function of $T_R$ at
(a) weak ($\alpha_{L_h}=\alpha_{L_c}=0.05$) and (b) strong  ($\alpha_{L_h}=\alpha_{L_c}=3$) left qubit-bath couplings,
with various right qubit-bath coupling strengthes.
The other parameters are given by $\varepsilon=1$, $\Delta=1$, $U=0.1$, $\omega_c=5$, $T_{L_h}=2$ and $T_{L_c}=0.2$.
}~\label{fig7}
\end{figure}
%%==========================================

Next, we try to find out separately that whether the strong coupling of left two thermal baths or the right bath exhibits the generation of the heat amplification.
 In Fig.~\ref{fig7}(a) with weak left qubit-bath couplings, it is interesting to find that the heat amplification occurs with strong coupling between the right qubit and corresponding bath.
While in Fig.~\ref{fig7}(b) with strong left qubit-bath couplings, a giant amplification factor is shown even at weak right qubit-bath coupling case ($e.g., \alpha_R=0.1$).
Hence, we conclude that strong coupling either from the left qubit-baths or right counterpart results in the heat amplification effect.
It should be noted these results are different from the previous work of nonequilibrium three-qubits system~\cite{cwang2018pra},
where the strong coupling from the thermal baths with moderate temperature (corresponding to $T_R$ in Fig.~\ref{fig1}(b)) is necessary to exhibit the amplification effect.

%no switch
% have stabilizer
% middle term dominate? check it
%using the NDTC to explain the HA.

\section{Conclusion}
To summarize, we study the quantum thermal transfer in the nonequilibrium two-qubits system,
by applying the nonequilibrium polaron-transformed Redfield equation combined with full counting statistics.
The heat current is unified in a broad coupling regime, which reduces to the Redfield and NIBA schemes in weak and strong interaction limits.
The negative differential thermal conductance is investigated within the two-terminal setup in Fig.~\ref{fig1}(a),
and the NDTC occurs with strong qubit-bath coupling and large temperature bias,
{\emph{which clearly answers the first question raised in the introduction}}.
The underlying mechanism is exploited based on the expression of the heat current at Eq.~(\ref{current-2}).
Moreover, by setting the weak left-qubit coupling, it is interesting to find that the NDTC sustains in strong right qubit-bath coupling regime.
{This clearly demonstrates that the partially strong qubit-bath interaction is sufficient to show the appearance of the NDTC.}
Next, we analyze the heat amplification effect within the three-terminal setup in Fig.~\ref{fig1}(b), where two qubits are asymmetrically coupled to three thermal baths.
A giant heat amplification factor is clearly observed in strong qubit-bath coupling regime with comparatively low temperature of the $R$th bath,
{\emph{which is the answer for the second question in the introduction.}}
It is tightly related with the turnover behavior of the heat current into the $R$th bath.
Moreover, it is discovered that either the strong interaction of the left qubit-bath interactions or the right counterpart can exhibit such
significant amplification effect.
We hope these findings can provide some insight {into designing the functional thermal transistor in coupled-qubits systems}.

\section{Acknowledgement}
H.L and C.W. are supported by the National Natural Science Foundation of China under Grant No. 11704093.
L.Q.W and J.R. acknowledge the support by the National Natural Science Foundation of China (No. 11775159), Natural Science Foundation of Shanghai (No. 18ZR1442800), and the National Youth 1000 Talents Program in China.

\appendix

\begin{comment}
\section{Calculation details}

\emph{Renormalization factor}

The spectral function is $J_v(\omega)=4\pi\sum_k|g_{k,v}|^2\delta(\omega-\omega_k)$.
The renormalization factor is
\begin{eqnarray}
\eta_v=\exp[-2\sum_k|g_{kv}/\omega_k|^2(2n_k+1)]=\exp[-\int^\infty_0d{\omega}\frac{J_v(\omega)}{\pi\omega^2}(n_v(\omega)+1/2)]\nonumber
\end{eqnarray}
The spectral function is specified as
$J_v(\omega)=\pi\alpha_v\omega^s\omega^{1-s}_{c,v}\exp(-|\omega|/\omega_{c,v})$ with $s=3$.
Hence,
\begin{eqnarray}
\eta_v=\exp\{\frac{\alpha_v}{2}[1-\frac{2}{(\beta_v\omega_{c,v})^2}\psi_1(\frac{1}{\beta_v\omega_{c,v}})]\}
\end{eqnarray}
with $\psi_1(x)=\sum^\infty_{n=0}\frac{1}{(n+x)^2}$.

\emph{Correlation function}

The correlation function is expressed as
\begin{eqnarray}
Q_v(t)&=&\int^\infty_0d{\omega}\frac{J_v(\omega)}{\pi\omega^2}[n_v(\omega)e^{i\omega{t}}+(1+n_v(\omega))e^{-i\omega{t}}]\nonumber\\
&=&\alpha_v[\frac{-1+(\omega_{c,v}t)^2}{(1+\omega^2_{c,v}t^2)^2}
+\frac{2}{(\beta_v\omega_{c,v})^2}\textrm{Re}\{\psi_1(\frac{1+i\omega_{c,v}t}{\beta_v\omega_{c,v}})\}]
-i\alpha_v\frac{2\omega_{c,v}t}{(1+\omega^2_{c,v}t^2)^2}\nonumber
\end{eqnarray}

\end{comment}

%\section{Full counting statistics of heat current}

%%==========================================
\begin{figure*}[tbp]
\begin{center}
\includegraphics[scale=0.5]{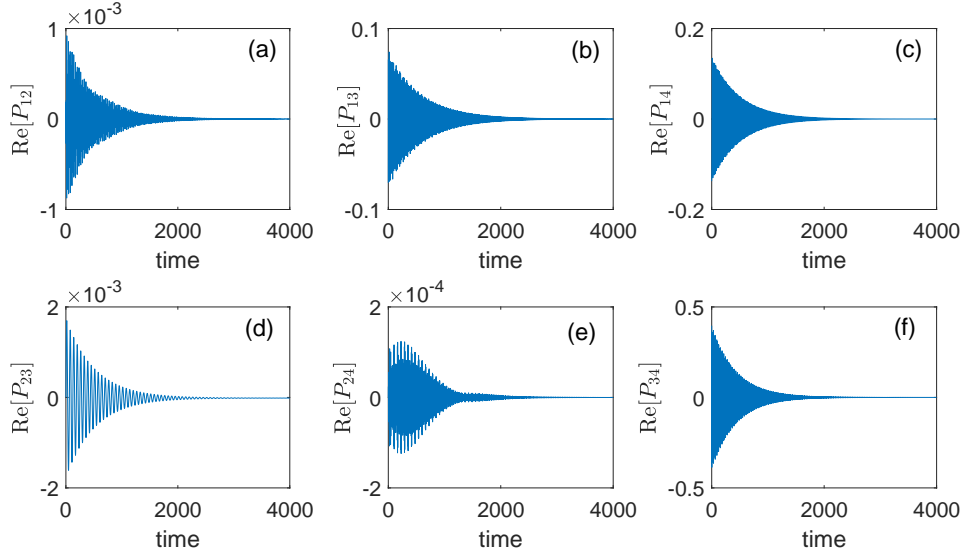}
%\vspace{-2.0cm}
\end{center}
\caption{(Color online) Dynamics of the real part of off-diagonal density matrix elements.
The system parameters are given by $\varepsilon=1$, $\Delta=1$, $U=0.1$, $\alpha=0.01$, $\omega_c=5$,
$T_L=1.5$ and $T_R=0.5$.
}~\label{append-a-fig1}
\end{figure*}
%%==========================================

\section{Steady state heat current within Redfield}

{The two-coupled spin-boson system is} expressed as $\hat{H}=\hat{H}_s+\sum_v(\hat{H}^v_b+\hat{V}_v)$,
where the qubits Hamiltonian is given by
\begin{eqnarray}
\hat{H}_s=\sum_v(\frac{\varepsilon_v}{2}\hat{\sigma}^v_z+\frac{\Delta_v}{2}\hat{\sigma}^v_x)+U\hat{\sigma}^L_z\hat{\sigma}^R_z,
\end{eqnarray}
the $v$th thermal bath is $\hat{H}^v_b=\sum_k\omega_k\hat{b}^{\dag}_{k,v}\hat{b}_{k,v}$,
and the system-bath interaction is
\begin{eqnarray}~\label{app-a-vv1}
\hat{V}_v=\sum_k(g_{kv}\hat{b}^{\dag}_{kv}+g^*_{kv}\hat{b}_{kv})\hat{\sigma}^v_z.
\end{eqnarray}
To count the energy flow into the right bath including full counting statistics, the total Hamiltonian is changed to
$\hat{H}(\chi)=e^{i\chi\hat{H}^R_b/2}\hat{H}e^{-i\chi\hat{H}^R_b/2}=\hat{H}_s+\sum_v(\hat{H}^v_b+\hat{V}^\chi_v)$,
where
\begin{eqnarray}
\hat{V}^\chi_v=\sum_k(g_{kv}e^{i\frac{\chi\omega_k}{2}\delta_{v,R}}\hat{b}^{\dag}_{kv}+g^*_{kv}e^{-i\frac{\chi\omega_k}{2}\delta_{v,R}}\hat{b}_{kv})\hat{\sigma}^v_z.
\end{eqnarray}
Considering the weak qubit-bath interaction, we perturb the system-bath interaction at Eq.~(\ref{app-a-vv1}) up to the second order.
Then, based on the Born-Markov approximation the Redfield equation is given by
\begin{eqnarray}
\frac{d\hat{\rho}_{\chi}(t)}{dt}&=&-i[\hat{H}_s,\hat{\rho}_{\chi}(t)]\\
&&-\sum_v\int^\infty_0{d\tau}\textrm{Tr}_b\{[\hat{V}^\chi_v,[\hat{V}^\chi_v(-\tau),\hat{\rho}_{\chi}(t){\otimes}\hat{\rho}_b]_\chi]_\chi\},\nonumber
\end{eqnarray}
where the commutating relation is $[\hat{A}_\chi,\hat{B}_\chi]_\chi=\hat{A}_\chi\hat{B}_\chi-\hat{B}_\chi\hat{A}_{-\chi}$.
In the eigen-basis $\{|\phi_n{\rangle}\}$ with $\hat{H}_s|\phi_n{\rangle}=E_n|\phi_n{\rangle}$,
the dynamical equation of the system density matrix element is given by
\begin{widetext}
\begin{eqnarray}
\frac{d\rho^{\chi}_{nn^{\prime}}}{dt}&=&-iE_{nn^{\prime}}\rho^{\chi}_{nn^{\prime}}
-\frac{1}{4}\sum_{v,m,m^{\prime}}J_v(E_{mm^{\prime}})n_v(E_{mm^{\prime}})\sigma^{nm}_{z,v}\sigma^{mm^{\prime}}_{z,v}\rho^{\chi}_{m^{\prime}n^{\prime}}\nonumber\\
&&-\frac{1}{4}\sum_{v,m,m^{\prime}}J_v(E_{mm^{\prime}})(1+n_v(E_{mm^{\prime}}))\sigma^{mm^{\prime}}_{z,v}\sigma^{m^{\prime}n^{\prime}}_{z,v}\rho^{\chi}_{nm}\nonumber\\
&&+\frac{1}{4}\sum_{v,m,m^{\prime}}J_v(E_{nm})n_v(E_{nm})e^{-iE_{nm}\chi\delta_{v,R}}\sigma^{nm}_{z,v}\sigma^{m^{\prime}n^{\prime}}_{z,v}\rho^{\chi}_{mm^{\prime}}\nonumber\\
&&+\frac{1}{4}\sum_{v,m,m^{\prime}}J_v(E_{m^{\prime}n^{\prime}})(1+n_v(E_{m^{\prime}n^{\prime}}))e^{iE_{m^{\prime}n^{\prime}}\chi\delta_{v,R}}
\sigma^{nm}_{z,v}\sigma^{m^{\prime}n^{\prime}}_{z,v}\rho^{\chi}_{mm^{\prime}}\nonumber
\end{eqnarray}
\end{widetext}
Moreover, from Fig.~\ref{append-a-fig1} it is known that off-diagonal elements becomes negligible at steady state.
Hence,  the Redfield equation after the long time evolution is reduced to
%\begin{widetext}
\begin{eqnarray}~\label{app-a-redfield}
\frac{d\rho^{\chi}_{nn}}{dt}&=&\sum_{m{\neq}n,v}\frac{J_v(E_{nm})}{2}n_v(E_{nm})e^{-iE_{nm}\chi\delta_{v,R}}\sigma^{nm}_{z,v}\sigma^{mn}_{z,v}\rho^{\chi}_{mm}\nonumber\\
&&-\sum_{m{\neq}n,v}\frac{J_v(E_{mn})}{2}n_v(E_{mn})\sigma^{nm}_{z,v}\sigma^{mn}_{z,v}\rho^{\chi}_{nn},
\end{eqnarray}
%\end{widetext}
Finally, the steady state heat flux is given by
\begin{eqnarray}~\label{current-redfield}
J_{\mathrm{Redfield}}=\sum_{n{\neq}m}\frac{E_{mn}}{2}J_R(E_{mn})(1+n_R(E_{mn}))\sigma^R_{nm}\sigma^R_{mn}P_{m}.\nonumber\\
%J_{Redfield}=\sum_{m}\Gamma_mP_{m}
\end{eqnarray}
%with the transition rate
%$\Gamma_m=\sum_nE_{mn}J_R(E_{mn})(1+n_R(E_{mn}))\sigma^{nm}_{z,R}\sigma^{mn}_{z,R}$.

%%==========================================
\begin{figure}[tbp]
\begin{center}
\includegraphics[scale=0.4]{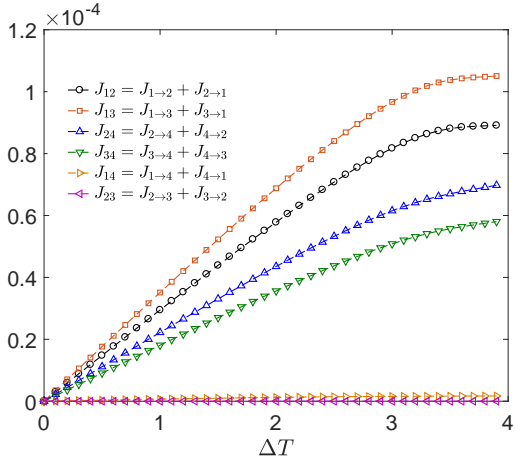}
%\vspace{-2.0cm}
\end{center}
\caption{(Color online) (a) Components of the heat current $J_{\mathrm{weak}}$ within the Redfield scheme at Eq.~(\ref{current-redfield})
and (b) the corresponding schematic illustration of heat transfer processes.
The bath temperatures are given by $T_L=T_0+{\Delta}T/2$ and $T_R=T_0-{\Delta}T/2$ with $T_0=2$.
The other system parameters are given by $\varepsilon=1$, $\Delta=1$, $U=0.1$, $\alpha=0.01$, $\omega_c=5$.
}~\label{append-a-fig2}
\end{figure}
%%==========================================

Then, we briefly discuss the absence of the NDTC within the Redfield scheme in Fig.~\ref{append-a-fig2}.
We first define the transition current from the population in the eigenspace $P_n$ to $P_m$ as
$J_{n{\rightarrow}m}=\frac{E_{mn}}{2}J_R(E_{mn})(1+n_R(E_{mn}))\sigma^R_{nm}\sigma^R_{mn}P_{m}$,
and the net transition current $J_{nm}=J_{n{\rightarrow}m}+J_{m{\rightarrow}n}$.
With the weak qubit-qubit coupling(e.g., $U=0.1$ in Fig.~\ref{append-a-fig2}), it is found that
$J_{14}$ and $J_{23}$ are negligible.
Thus, the heat current can be simplified as
\begin{eqnarray}~\label{j-weak-1}
J_{\mathrm{Redfield}}{\approx}J_{12}+J_{13}+J_{24}+J_{34}.
\end{eqnarray}
By increasing the temperature bias, the contributing net currents at Eq.~(\ref{j-weak-1}) show monotonic increase,
resulting in the enhancement of the $J_{\mathrm{Redfield}}$.
Hence, $J_{\mathrm{Redfield}}$ will not exhibit the NDTC.

%%==========================================
%\begin{figure*}[tbp]
%\begin{center}
%\includegraphics[scale=0.6]{append_A_Redfield_ms.eps}
%\vspace{-2.0cm}
%\end{center}
%\caption{(Color online) Normalized heat current $I/I_{max}(\alpha)$ with various (a) qubit splitting energies $\varepsilon$ and (b) qubit-qubit repulsion strengthes $U$, steady state populations with (c) $\varepsilon=0$ and (e) $U=0.8$, transition rates with (d) $\varepsilon=0$ and (f) $U=0.8$,
%by tuning temperature bias $\Delta{T}$ with strong qubit-bath coupling $\alpha=5$.
%}~\label{append-a-fig2}
%\end{figure*}

%%==========================================
\section{Steady state heat current within NIBA}
Following the similar treatment in Sec. II (B), after the polaron transformation the system Hamiltonian is given by
\begin{eqnarray}
\hat{H}^{\prime}_s=\frac{\varepsilon_L}{2}\hat{\sigma}^L_z+\frac{\varepsilon_R}{2}\hat{\sigma}^R_z+U\hat{\sigma}^L_z\hat{\sigma}^R_z,
\end{eqnarray}
and the system-bath interaction is
\begin{eqnarray}
V_v=\hat{\sigma}^v_x\cos\hat{B}_v+\hat{\sigma}^v_y\sin\hat{B}_v,
\end{eqnarray}
with the collective phonon momentum $\hat{B}_v=2i\sum_k(g_{k,v}/\omega_k\hat{b}^{\dag}_{k,v}-H.c.)$.
By applying the full counting statistics, the counting field dependent qubit-bath interaction is changed to
$V^\chi_v=\hat{\sigma}^v_x\cos\hat{B}^\chi_v+\hat{\sigma}^v_y\sin\hat{B}^\chi_v$,
with $\hat{B}^\chi_v=2i\sum_k(\frac{g_{k,v}}{\omega_k}e^{\frac{i\chi}{2}\omega_k\delta_{v,R}}\hat{b}^{\dag}_{k,v}-H.c.)$.
Consequently, the quantum master equation is given by
\begin{eqnarray}~\label{app-b-NIBA}
\frac{d\hat{\rho}_s(t)}{dt}&=&-i[\hat{H}^{\prime}_s,\hat{\rho}_s(t)]
+\sum_{v=L,R}\int^\infty_0d{\tau}\\
&&{\times}\{D_v(\chi,-\tau)[\hat{\sigma}^v_+\hat{\rho}_{\chi}(t)\hat{\sigma}^v_-(-\tau)+\hat{\sigma}^v_-\hat{\rho}_{\chi}(t)\hat{\sigma}^v_+(-\tau)]\nonumber\\
&&+D_v(\chi,\tau)[\hat{\sigma}^v_+(-\tau)\hat{\rho}_{\chi}(t)\hat{\sigma}^v_-+\hat{\sigma}^v_-(-\tau)\hat{\rho}_{\chi}(t)\hat{\sigma}^v_+]\nonumber\\
&&-D_v(\tau)[\hat{\sigma}^v_+\hat{\sigma}^v_-(-\tau)\hat{\rho}_{\chi}(t)+\hat{\sigma}^v_-\hat{\sigma}^v_+(-\tau)\hat{\rho}_{\chi}(t)]\nonumber\\
&&-D_v(-\tau)[\hat{\rho}_{\chi}(t)\hat{\sigma}^v_+(-\tau)\hat{\sigma}^v_-+\hat{\rho}_{\chi}(t)\hat{\sigma}^v_-(-\tau)\hat{\sigma}^v_+]\}\nonumber
\end{eqnarray}
where the transition correlation function is
\begin{eqnarray}
D_v(\chi,\pm\tau)&=&(\frac{\Delta_v}{2})^2\exp[-4\sum_{k}|\frac{g_{kv}}{\omega_k}|^2(2n_k+1)\nonumber\\
&&+Q_v(\tau-\chi\delta_{v,R})].
\end{eqnarray}
Specifically, the population dynamics in local basis
$\{|\phi_1{\rangle}=|\uparrow\uparrow{\rangle},|\phi_2{\rangle}=|\uparrow\downarrow{\rangle},|\phi_3{\rangle}=|\downarrow\uparrow{\rangle},|\phi_4{\rangle}=|\downarrow\downarrow{\rangle},\}$ are
\begin{eqnarray}
\frac{dP^{\chi}_1(t)}{dt}&=&\kappa^{31}_LP^{\chi}_3(t)+\kappa^{21}_R({\chi})P^{\chi}_2(t)-(\kappa^{13}_L+\kappa^{12}_R)P^{\chi}_1(t)\nonumber\\
\frac{dP^{\chi}_2(t)}{dt}&=&\kappa^{42}_LP^{\chi}_4(t)+\kappa^{12}_R({\chi})P^{\chi}_1(t)
-(\kappa^{24}_L+\kappa^{21}_R)P^{\chi}_2(t)\nonumber\\
\frac{dP^{\chi}_3(t)}{dt}&=&\kappa^{13}_LP^{\chi}_1(t)+\kappa^{43}_R({\chi})P^{\chi}_4(t)
-(\kappa^{31}_L+\kappa^{34}_R)P^{\chi}_3(t)\nonumber\\
\frac{dP^{\chi}_4(t)}{dt}&=&\kappa^{24}_LP^{\chi}_2(t)+\kappa^{34}_R({\chi})P^{\chi}_3(t)
-(\kappa^{42}_L+\kappa^{43}_R)P^{\chi}_4(t)\nonumber
\end{eqnarray}
where $P^{\chi}_i={\langle}\phi_i|\hat{\rho}_{\chi}|\phi_i{\rangle}$
the counting field dependent transition rates are $\kappa^{ij}_v({\chi})=e^{iE_{ij}\chi}\kappa^{ij}_v$,
and the standard transition rates are
\begin{eqnarray}~\label{app-b-rate}
\kappa^{ij}_v&=&\int^\infty_{-\infty}d{\tau}D_v(\tau)e^{iE_{ij}\tau}.
\end{eqnarray}
Consequently, the heat current is expressed as
%\begin{eqnarray}
%J&=&E_{21}[\kappa_R(E_{21})P_2-\kappa_R(-E_{21})P_1]
%+E_{43}[\kappa_R(E_{43})P_4-\kappa_R(-E_{43})P_3].
%\end{eqnarray}
\begin{eqnarray}~\label{current-niba}
I_{\mathrm{NIBA}}&=&E_{12}[\kappa^{12}_RP_1-\kappa^{21}_RP_2]
-E_{34}[\kappa^{43}_RP_4-\kappa^{34}_RP_3].\nonumber\\
\end{eqnarray}
Under the condition $\varepsilon_v=\varepsilon$,
the steady state populations in absence of the counting field parameter are given by
\begin{eqnarray}
P_1&=&(\kappa^{21}_L\kappa^{21}_R\kappa^{42}_R+\kappa^{42}_L\kappa^{21}_R\kappa^{24}_R+\kappa^{21}_L\kappa^{42}_L\kappa^{21}_R+\kappa^{21}_L\kappa^{24}_L\kappa^{42}_R)/\mathcal{A}\nonumber\\
P_2&=&[\kappa^{21}_L\kappa^{12}_R\kappa^{42}_R+(\kappa^{12}_L+\kappa^{12}_R)\kappa^{42}_L\kappa^{24}_R+\kappa^{21}_L\kappa^{42}_L\kappa^{12}_R]/\mathcal{A},\nonumber\\
P_3&=&[\kappa^{21}_R\kappa^{12}_L\kappa^{42}_L+(\kappa^{12}_L+\kappa^{12}_R)\kappa^{42}_R\kappa^{24}_L+\kappa^{21}_R\kappa^{42}_R\kappa^{12}_L]/\mathcal{A}\nonumber\\
P_4&=&[\kappa^{21}_L\kappa^{24}_L\kappa^{12}_R+\kappa^{21}_R\kappa^{24}_R\kappa^{12}_L+(\kappa^{12}_L+\kappa^{12}_R)\kappa^{24}_L\kappa^{24}_R]/\mathcal{A},
\end{eqnarray}
where the coefficient $\mathcal{A}$ is
\begin{eqnarray}
\mathcal{A}&=&[\kappa^{12}_L\kappa^{42}_L\kappa^{21}_R+\kappa^{21}_L\kappa^{24}_L\kappa^{42}_R+\kappa^{21}_L\kappa^{42}_L\kappa^{21}_R
+\kappa^{21}_L\kappa^{42}_L\kappa^{12}_R\nonumber\\
&&+\kappa^{21}_L\kappa^{24}_L\kappa^{12}_R+(\kappa^{12}_L+\kappa^{12}_R)(\kappa^{24}_L\kappa^{42}_R+\kappa^{24}_L\kappa^{24}_R/2)]\nonumber\\
&&+[L{\leftrightarrow}R].\nonumber
\end{eqnarray}
And the current is simplified as
\begin{eqnarray}~\label{I_strong_1}
I_{\mathrm{NIBA}}&=&\frac{4U}{\mathcal{A}}(\kappa^{24}_L\kappa^{43}_R\kappa^{31}_L\kappa^{12}_R-\kappa^{21}_R\kappa^{13}_L\kappa^{34}_R\kappa^{42}_L).
\end{eqnarray}

%%==========================================
\begin{figure}[tbp]
\begin{center}
\includegraphics[scale=0.4]{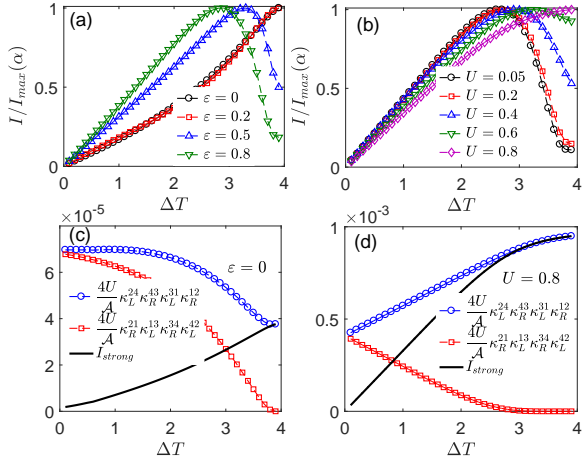}
%\vspace{-2.0cm}
\end{center}
\caption{(Color online) Normalized heat current $I/I_{max}(\alpha)$ with various (a) qubit splitting energies $\varepsilon$ and (b) qubit-qubit repulsion strengthes $U$, loop currents and $I_{strong}$ with (c) $\varepsilon=0$ and (d) $U=0.8$,
by tuning temperature bias $\Delta{T}$ with strong qubit-bath coupling $\alpha=5$.
The other parameters are the same as in Fig.~\ref{fig3}.
}~\label{append-b-fig1}
\end{figure}

{%\emph{Effects of intra-qubit energy bias and inter-qubit coupling}
From Fig.~\ref{fig3}, it is known that the NDTC occurs with finite qubit splitting energy with strong qubit-bath coupling.
Here, we analyze the influence of the splitting energy on the steady state heat current in Fig.~\ref{append-b-fig1}(a).
At resonance ($\varepsilon=0$), the Normalized heat current shows monotonic enhancement by increasing temperature bias (${\Delta}T$) without the NDTC.
From the Eq.~(\ref{I_strong_1}), the heat current is contributed by two loop currents
$\frac{4U}{\mathcal{A}}\kappa^{24}_L\kappa^{43}_R\kappa^{31}_L\kappa^{12}_R$ and $\frac{4U}{\mathcal{A}}\kappa^{21}_R\kappa^{13}_L\kappa^{34}_R\kappa^{42}_L$.
In Fig.~\ref{append-b-fig1}, it is found that the second component  $\frac{4U}{\mathcal{A}}\kappa^{21}_R\kappa^{13}_L\kappa^{34}_R\kappa^{42}_L$
is strongly suppressed by increasing ${\Delta}T$,
which enhances the difference between two loop currents.
Hence, $I_{\mathrm{NIBA}}$ exhibits monotonic behavior.
Next by turning on $\varepsilon$ (e.g., $\varepsilon=0.5,0.8$), it is found the NDTC finally occurs.
Therefore, we conclude that qubit splitting energy is necessary to show the NDTC in the nonequilibrium two-qubit system.

Then, we study the effect of quibt-qubit interaction on the heat current in Fig.~\ref{append-b-fig1}(b).
It is found that by increasing the repulsion strength $U$, the behavior of NDTC is suppressed accordingly,
and completely vanished with strong repulsion strength(e.g., $U=0.8$).
With strong qubit-qubit repulsion, the current $\frac{4U}{\mathcal{A}}\kappa^{24}_L\kappa^{43}_R\kappa^{31}_L\kappa^{12}_R$  is apparently enhanced by increasing ${\Delta}T$,
whereas $\frac{4U}{\mathcal{A}}\kappa^{21}_R\kappa^{13}_L\kappa^{34}_R\kappa^{42}_L$ shows decrease in Fig.~\ref{append-b-fig1}(d).
It finally contributes to the monotonic increase of $I_{\mathrm{NIBA}}$.
}

%%==========================================
\begin{figure}[tbp]
\begin{center}
%\vspace{-1.0cm}
\includegraphics[scale=0.45]{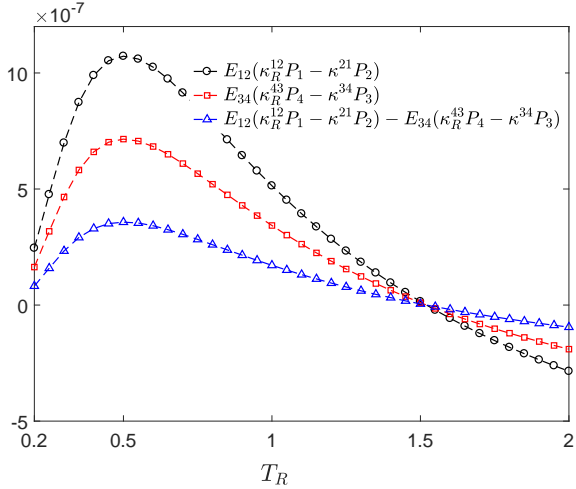}
%\vspace{-2.0cm}
\end{center}
\caption{(Color online) Heat current and components at Eq.~(\ref{app-c-IR}) within the three-terminal setup
with strong qubit-bath coupling ($\alpha=5$) as a function of $T_R$.
The other parameters are $\varepsilon=1$, $\Delta=1$, $U=0.1$, $\omega_c=5$, $T_{L_h}=2$ and $T_{L_c}=0.2$.
}~\label{append-c-fig1}
\end{figure}
%%==========================================
\section{Heat current in two-qubits asymmetrically coupled to three thermal baths}

The two qubits system asymmetrically coupled to three thermal baths,  which is shown in Fig.~\ref{fig1}(b), is modeled as
\begin{eqnarray}
\hat{H}=\hat{H}_s+\sum_{v=L_h,L_c,R}(\hat{H}^v_b+\hat{V}_v).
\end{eqnarray}
The system Hamiltonian is given by
\begin{eqnarray}
\hat{H}_s=U\hat{\sigma}^L_z\hat{\sigma}^R_z+\sum_{v=L,R}(\frac{\varepsilon_v}{2}\hat{\sigma}^v_z+\frac{\Delta_v}{2}\hat{\sigma}^v_x).
\end{eqnarray}
The $v$th thermal bath is described as $\hat{H}^{v}_b=\sum_k\omega_k\hat{b}^{\dag}_{k,v}\hat{b}_{k,v}$.
The $L_u$th qubit-bath interaction is given by
\begin{eqnarray}
\hat{V}_{L_u}=\hat{\sigma}^L_z\sum_{k}(g_{k,L_u}\hat{b}^{\dag}_{k,L_u}+g^*_{k,L_u}\hat{b}_{k,L_u}),
\end{eqnarray}
and the $R$th qubit-bath interaction is
\begin{eqnarray}
\hat{V}_R=\hat{\sigma}^R_z\sum_{k}(g_{k,R}\hat{b}^{\dag}_{k,R}+g^*_{k,R}\hat{b}_{k,R}).
\end{eqnarray}
Then, we apply the full counting statistics to count the heat flow into the $v$th thermal bath as
\begin{eqnarray}
\hat{H}(\{\chi\})&=&\exp[\frac{i}{2}(\sum_v\chi_{v}\hat{H}^{v}_b)]\hat{H}\exp[-\frac{i}{2}(\sum_v\chi_{v}\hat{H}^{v}_b)]\nonumber\\
&=&\hat{H}_s+\sum_{v=L_h,L_c,R}[\hat{H}^v_b+\hat{V}_v(\chi_v)],
\end{eqnarray}
where the counting parameter set is $\{\chi\}=(\chi_{L_h},\chi_{L_c},\chi_R)$,
$\hat{V}_{L_u}(\chi_{L_u})=\hat{\sigma}^L_z\sum_{k}(g_{k,L_u}e^{\frac{i\omega_k\chi_{L_u}}{2}}\hat{b}^{\dag}_{k,L_u}
+H.c.)$
and
$\hat{V}_{R}(\chi_{R})=\hat{\sigma}^R_z\sum_{k}(g_{k,R}e^{\frac{i\omega_k\chi_{R}}{2}}\hat{b}^{\dag}_{k,R}+H.c.)$.

Similar to the treatment at Eq.~(\ref{fcs-hv}), we apply the canonical transformation to obtain the modified Hamiltonian as
\begin{eqnarray}
\hat{H}^{\prime}(\{\chi\})=\hat{U}^{\dag}_{\{\chi\}}\hat{H}(\{\chi\})\hat{U}_{\{\chi\}},
\end{eqnarray}
where the unitary operator is given by
$\hat{U}_{\{\chi\}}=\exp\{i[(\sum_u\hat{B}_{L_u}(\chi_{L_u}))\hat{\sigma}^L_z+\hat{B}_R(\chi_R)\hat{\sigma}^R_z]$,
with the collective phononic momentum
$\hat{B}_v(\chi_v)=2i\sum_k(g_{kv}e^{\frac{i\omega_k\chi_v}{2}}/\omega_k\hat{b}^{\dag}_{kv}-H.c.)$.
The modified Hamiltonian is specified as
$\hat{H}^{\prime}(\{\chi\})=\hat{H}^{\prime}_s+\hat{V}^{\prime}(\{\chi\})+\sum_{v}\hat{H}^v_b$,
where the transformed qubits Hamiltonian is
\begin{eqnarray}
\hat{H}^{\prime}_s=\sum_{v=L,R}(\frac{\varepsilon_v}{2}\hat{\sigma}^v_z+\frac{\eta_v\Delta_v}{2}\hat{\sigma}^v_x)+U\hat{\sigma}^L_z\hat{\sigma}^R_z,
\end{eqnarray}
with the renormalization factors
\begin{eqnarray}
\eta_L&=&\exp[-2\sum_{k;u=h,c}|g_{k,L_u}/\omega_k|^2(2n_{k,L_u}+1)],\\
\eta_R&=&\exp[-2\sum_k|g_{kR}/\omega_k|^2(2n_{k,R}+1)],\nonumber
\end{eqnarray}
The system-bath interaction is given by
\begin{eqnarray}~\label{app-vv}
\hat{V}^{\prime}(\{\chi\})&=&[\hat{\sigma}^L_x(\cos\sum_{u}\hat{B}_{L_u}(\chi_{L_u})-\eta_L)\\
&&+\hat{\sigma}^L_y\sin\sum_{u}\hat{B}_{L_u}(\chi_{L_u})]\nonumber\\
&&+[\hat{\sigma}^R_x(\cos\hat{B_R}(\chi_{R})-\eta_R)+\hat{\sigma}^R_y\sin\hat{B}_R(\chi_{R})].\nonumber
\end{eqnarray}

Then, based on the Born-Markov approximation and perturbing the system-bath interaction  up to the second order,
we obtain the quantum master equation as
\begin{widetext}
\begin{eqnarray}~\label{app-c-qme}
\frac{d\hat{\rho}_{\{\chi\}}(t)}{dt}&=&-i[\hat{H}^{\prime}_s,\hat{\rho}_{\{\chi\}}(t)]+\sum_{v=L,R;a=x,y;\omega,\omega^{\prime}}
[(\Gamma^v_{a,-}(\{\chi\},\omega)+\Gamma^v_{a,+}(\{\chi\},\omega^{\prime}))\hat{P}^v_a(\omega^{\prime})\hat{\rho}_{\{\chi\}}(t)\hat{P}^v_a(\omega)\nonumber\\
&&-(\Gamma^v_{a,+}(\omega)\hat{P}^v_a(\omega^{\prime})\hat{P}^v_a(\omega)\hat{\rho}_{\{\chi\}}(t)+H.c.)],\nonumber
\end{eqnarray}
\end{widetext}
where the transition rates are
\begin{eqnarray}~\label{app-c-rate}
\Gamma^L_{a,\pm}(\{\chi\},\omega)&=&\int^{\infty}_0d{\tau}C^L_a(\chi_{L_h},\chi_{L_c},\pm\tau)e^{i\omega\tau}\\
\Gamma^R_{a,\pm}(\{\chi\},\omega)&=&\int^{\infty}_0d{\tau}C^R_a(\chi_{R},\pm\tau)e^{i\omega\tau}.\nonumber
\end{eqnarray}
and the counting-field dependent correlation functions are
\begin{eqnarray}
C^L_x(\chi_{L_h},\chi_{L_c},\tau)&=&(\frac{\eta_L\Delta_L}{2})^2[\cosh(\sum_ue^{Q_{L_u}(\tau-\chi_{L_u})})-1]\nonumber\\
C^L_y(\chi_{L_h},\chi_{L_c},\tau)&=&(\frac{\eta_L\Delta_L}{2})^2\sinh(\sum_ue^{Q_{L_u}(\tau-\chi_{L_u})}),\nonumber\\
C^R_x(\chi_R,\tau)&=&(\frac{\eta_R\Delta_R}{2})^2[\cosh(e^{Q_R(\tau-\chi_R)})-1]\nonumber\\
C^R_y(\chi_R,\tau)&=&(\frac{\eta_R\Delta_R}{2})^2\sinh(e^{Q_R(\tau-\chi_R)}),
\end{eqnarray}
where the correlation phase is
\begin{eqnarray}
Q_v(\tau)=4\sum_{k}|\frac{g_{kv}}{\omega_k}|^2[\cos\omega_k\tau(2n_{k,v}+1)-i\sin\omega_k\tau].\nonumber\\
\end{eqnarray}
Finally, the cumulant generating function after a long time evolution is given by
$\mathcal{Z}(\{\chi\})=\lim_{t{\rightarrow}\infty}\frac{1}{t}\ln\textrm{Tr}\{\hat{\rho}_{\{\chi\}}(t)\}$,
and the steady state heat current into the $v$th bath is
\begin{eqnarray}
I_v={\frac{{\partial}\mathcal{Z}(\{\chi\})}{{\partial}(i\chi_v)}}{|}_{\{\chi\}=0}.
\end{eqnarray}

Comparing the dynamical equation of qubits density matrix at Eq.~(\ref{app-c-qme}) with two-terminal case at Eq.~(\ref{ptre-chi}),
it is interesting to find that two equations are almost identical except expressions of the transition rates related with left bath(s)
$\Gamma^L_{a,\pm}(\chi,\omega)$ at Eq.~(\ref{twoterminal-rate}) and $\Gamma^{L}_{a,\pm}(\{\chi\},\omega)$ at Eq.~(\ref{app-c-rate}).
Hence, considering strong qubit-bath interaction, the heat flux into the $R$th bath can be expressed as
\begin{eqnarray}~\label{app-c-IR}
I^R_{\mathrm{NIBA}}&=&E_{12}(\kappa^{12}_RP_1-\kappa^{21}_RP_2)
-E_{34}(\kappa^{43}_RP_4-\kappa^{34}_RP_3),\nonumber\\
\end{eqnarray}
where $\kappa^{ij}_R$ is identical with the counterpart at Eq.~(\ref{app-b-rate}).

{Then, we briefly discuss the nonmonotonic behavior of $I_R$ with strong qubit-bath coupling in Fig.~\ref{fig6}(b).
From Eq.~(\ref{app-c-IR}), it is known that $I_R$ is composed by two components
$E_{12}(\kappa^{12}_RP_1-\kappa^{21}_RP_2)$ and $E_{34}(\kappa^{43}_RP_4-\kappa^{34}_RP_3)$.
Both show nonmonotonic features by increasing temperature $T_R$, shown in Fig.~\ref{append-c-fig1}.
In particular, the gap between two current components is first enhanced in the regime $T_R{\in}(0.2,0.5)$,
and then decreases in the regime $T_R{\in}(0.5,1.5)$.
This directly results in the turnover behavior of $I_R$ by tuning the temperature $T_R$.
}

\end{document}